\documentclass[sigconf]{acmart}

\AtBeginDocument{%
  }

\copyrightyear{2026}
\acmYear{2026}
\setcopyright{cc}
\setcctype{by-nc-nd}
\acmConference[FORGE '26]{2026 IEEE/ACM Third International Conference on AI Foundation Models and Software Engineering}{April 12--13, 2026}{Rio de Janeiro, Brazil}
\acmBooktitle{2026 IEEE/ACM Third International Conference on AI Foundation Models and Software Engineering (FORGE '26), April 12--13, 2026, Rio de Janeiro, Brazil}
\acmPrice{}
\acmDOI{10.1145/3793655.3793725}
\acmISBN{979-8-4007-2477-0/2026/04}

\usepackage{tikz}
\usepackage{caption}
\usepackage{blindtext}
\usepackage{tcolorbox}
\usepackage[final]{pdfpages}
\usepackage{lipsum,multicol}
\usepackage{xcolor}
\usepackage{tikz}
\usepackage{listings}
\usepackage{enumitem}
\usepackage{hyperref}
\usepackage{amsfonts}
\usepackage{wrapfig}
\usepackage{subcaption} 
\usepackage{adjustbox}
\usepackage{colortbl}
\usepackage{fancybox}
\usepackage{multirow}
\usepackage[normalem]{ulem}
\useunder{\uline}{\ul}{}
\usepackage{enumitem}
\usepackage{amsmath}
\usepackage{booktabs} 
\usepackage{array}

\usepackage{tabularx}

\newcolumntype{L}{>{\raggedright\arraybackslash}X}

\newcommand{\eg}{\textit{e.g.,}\xspace}

\newtcolorbox{boxK}{
    fontupper = \small,
    sharpish corners, %
    boxrule = 0pt,
    toprule = 0pt, %
}

\newcommand{\figref}[1]{Fig.~\ref{#1}\xspace}

\newcommand*\circled[1]{\tikz[baseline=(char.base)]{
            \node[shape=circle,draw,inner sep=0.5pt] (char) {#1};}}

\definecolor{gradientplum}{RGB}{196,115,156}

\definecolor{gradientplum}{RGB}{196,115,156}

\newcommand{\KCH}{\textit{KCH}\xspace}
\newcommand{\AST}{\textit{AST}\xspace}
\newcommand{\KCHs}{\textit{KCHs}\xspace}
\newcommand{\ASTs}{\textit{ASTs}\xspace}

\newcommand{\approptoinn}[2]{\mathrel{\vcenter{
  \offinterlineskip\halign{\hfil$##$\cr
    #1\propto\cr\noalign{\kern2pt}#1\sim\cr\noalign{\kern-2pt}}}}}

\definecolor{gradientplum}{RGB}{196,115,156}
\newcommand\repository[1]{\textcolor{gradientplum}{\href{#1}{repository}}}

\begin{document}
\title{ 
Detecting and Correcting Hallucinations in LLM-Generated Code via Deterministic AST Analysis
}

\settopmatter{authorsperrow=4}
\author{Dipin Khati}
\email{dkhati@wm.edu}
\affiliation{%
  \institution{William \& Mary}
  \city{Williamsburg}
  \state{Virginia}
  \country{USA}
}

\author{Daniel Rodriguez-Cardenas}
\email{dhrodriguezcar@wm.edu}
\affiliation{%
  \institution{William \& Mary}
  \city{Williamsburg}
  \state{Virginia}
  \country{USA}
}

\author{Paul Pantzer}
\email{pwpantzer@wm.edu}
\affiliation{%
  \institution{William \& Mary}
  \city{Williamsburg}
  \state{Virginia}
  \country{USA}
}

\author{Denys Poshyvanyk}
\email{dposhyvanyk@wm.edu}
\affiliation{%
  \institution{William \& Mary}
  \city{Williamsburg}
  \state{Virginia}
  \country{USA}
}

\renewcommand{\shortauthors}{Trovato et al.}

\begin{abstract}
Large Language Models (LLMs) for code generation boost productivity but frequently introduce Knowledge Conflicting Hallucinations (KCHs), subtle, semantic errors, such as non-existent API parameters, that evade linters and cause runtime failures. Existing mitigations like constrained decoding or non-deterministic LLM-in-the-loop repair are often unreliable for these errors. This paper investigates whether a deterministic, static-analysis framework can reliably detect \textit{and} auto-correct KCHs. We propose a post-processing framework that parses generated code into an Abstract Syntax Tree (AST) and validate it against a dynamically-generated Knowledge Base (KB) built via library introspection. This non-executing approach uses deterministic rules to find and fix both API and identifier-level conflicts. On a manually-curated dataset of 200 Python snippets, our framework detected KCHs with 100\% precision and 87.6\% recall (0.934 F1-score), and successfully auto-corrected 77.0\% of all identified hallucinations. Our findings demonstrate that this deterministic post-processing approach is a viable and reliable alternative to probabilistic repair, offering a clear path toward trustworthy code generation.
\end{abstract}




\maketitle

\section{Introduction}\label{sec:introduction}

LLMs have fundamentally reshaped the software development life cycle. For millions of developers, tools like GitHub Copilot now integrate deeply into daily workflow, acting as collaborative partners that demonstrably boost productivity by generating complex functions from a simple natural language comment \cite{peng2023impactaideveloperproductivity}. However, this new productivity introduces a subtle and frustrating class of errors that can erode developer trust \cite{khatiTrust}. Empirical studies on LLM-generated code confirm that models frequently produce outputs that appear syntactically correct but fail at runtime \cite{wang2025understandingcharacteristicscodegeneration, tambon2024bugslargelanguagemodels}. For example, when generating a pandas script, a model may call \texttt{pd.\allowbreak read\_exel('data.csv')}, which both invents a function name and mismatches the intended file type \cite{liu2024exploringevaluatinghallucinationsllmpowered}. This failure often presents as a block of code that looks clean, seems idiomatic, and even follows local variable naming conventions. But when executed, the program crashes, citing an ``unexpected keyword argument'' or an ``undefined variable.'' A linter, for instance, will not catch these errors; it’s not simple typo. These errors, in fact, represent a frequent and well-documented LLM bug pattern \cite{wang2025understandingcharacteristicscodegeneration, tambon2024bugslargelanguagemodels, wen2025fixingfunctionlevelcodegeneration}, a failure that reveals a deep structural error also known as hallucination.

This deep structural error, or hallucination, falls into a category that recent work defines as \KCH \cite{liu2024exploringevaluatinghallucinationsllmpowered}. This specific failure occurs when generated code flat-out contradicts the established, factual knowledge of a programming language or its libraries. For example, a model might seem to understand a library but confidently invents a parameter name that remains semantically plausible but factually incorrect. These errors represent the worst of both worlds. They do not constitute syntax errors; an invented parameter, for example, often results in perfectly valid syntax. Instead, they constitute deep semantic errors that only reveal themselves at runtime, making them incredibly costly to find in a complex application. This \KCH category encompasses two major problems: API Knowledge Conflicts, which involve using a deprecated function or non-existent parameter, and Identifier Knowledge Conflicts, where the model misuses a variable it defined itself, such as defining max\_len\_str but later calling max\_len\_len\_str \cite{liu2024exploringevaluatinghallucinationsllmpowered}.

As this kind of subtly broken, ``plausible-but-wrong'' code quietly filters into production repositories \cite{lee2025hallucinationcodegenerationllms}, the research community has scrambled to tackle the problem. However, the most prominent academic solutions focus on prevention rather than cure, primarily through constrained decoding techniques. Tools such as PICARD \cite{scholak-etal-2021-picard} and Synchromesh \cite{poesia2022synchromeshreliablecodegeneration} show real promise, forcing the LLM to only output tokens that fit a strict formal grammar. They do a solid job of ensuring syntactically valid code, eliminating errors such as missing brackets. But they remain blind to our specific problem. A function call with an invented parameter remains syntactically perfect, so these tools allow its generation, even when it contains deep semantic errors.

Other approaches pivot from prevention to repair. The most common repair method relies on an ``LLM-in-the-loop,'' where a static analysis tool finds an error and feeds its report back to the LLM to "fix the problem" \cite{blyth2025staticanalysisfeedbackloop, raviLLMLoop}. This approach, however, amounts to a non-deterministic gamble; it relies on the very model that hallucinated to suddenly understand its own subtle, factual error. A more recent technique, ``Structural Trimming,'' also uses a post-processing AST parser but focuses explicitly on deletion \cite{zhang2025cutting}. This method identifies the ``anchor'' of a hallucination and simply prunes the offending code block. While this presents a powerful idea for mitigating security vulnerabilities, it fails to solve the developer's core problem. It leaves behind an incomplete function. It does not resolve the error; it merely removes it.

This clear gap in the research leads us to our core research questions:

\begin{enumerate}[label=\textbf{RQ$_{\arabic*}$}, ref=\textbf{RQ$_{\arabic*}$}, labelindent=2em]\setlength{\itemsep}{0.2em}
    \item \label{rq:explore} Can we use classic static analysis techniques to deterministically detect the semantic, knowledge-conflicting hallucinations that LLMs produce?

    \item \label{rq:beyond} Can such a technique go beyond detection and auto-correct these specific errors in a post-processing step?
\end{enumerate}

To this end, we propose a lightweight, deterministic parser that resolves these hallucinations by trusting the code's concrete structure. The framework functions by parsing the generated code into its \AST, which provides a verifiable, logical map of the program. This map allows \textbf{our framework} to deterministically cross-reference every variable used against every variable defined in its scope, and to validate every API call against a "ground-truth" file of library signatures. For a developer, this means that frustrating runtime-crashing errors can be caught and fixed *before* execution, saving significant debugging time and rebuilding trust in the generated code. 

We evaluated our framework on a curated dataset of 200 samples designed to test \KCHs in multiple target libraries. The framework achieves a precision of 100\% in detecting hallucinations, indicating zero false positives, and a fix accuracy of 77.0\% in automatically correcting them. These results demonstrate that a static, deterministic approach can effectively detect and fix KCH to promote trustworthy code generation without relying on probabilistic post-hoc filtering.

While prior post-processing tools have used ASTs for pruning or formatting, our work is, to our knowledge, the first to apply a deterministic AST–KB combination for hallucination correction. Our approach offers a reliable, interpretable, and non-probabilistic mechanism for mitigating \KCHs in code produced by large language models. Additionally, our tool is standalone and can also be used on any codebase to scan for \KCHs, even on human-written code. All data, code, and experimental configurations are publicly available in our replication package~\cite{repo}.

\section{Methodology}
\label{sec:approach}

We introduce a deterministic, post-generation framework to detect and correct \KCHs in code. Our design is entirely static and non-executing: code is analyzed via \ASTs, validated against a versioned KB built from the libraries themselves, and repaired through localized \AST edits. This yields reproducible decisions and interpretable corrections without running the code.

\begin{figure*}
    \centering
    \includegraphics[width=\linewidth]{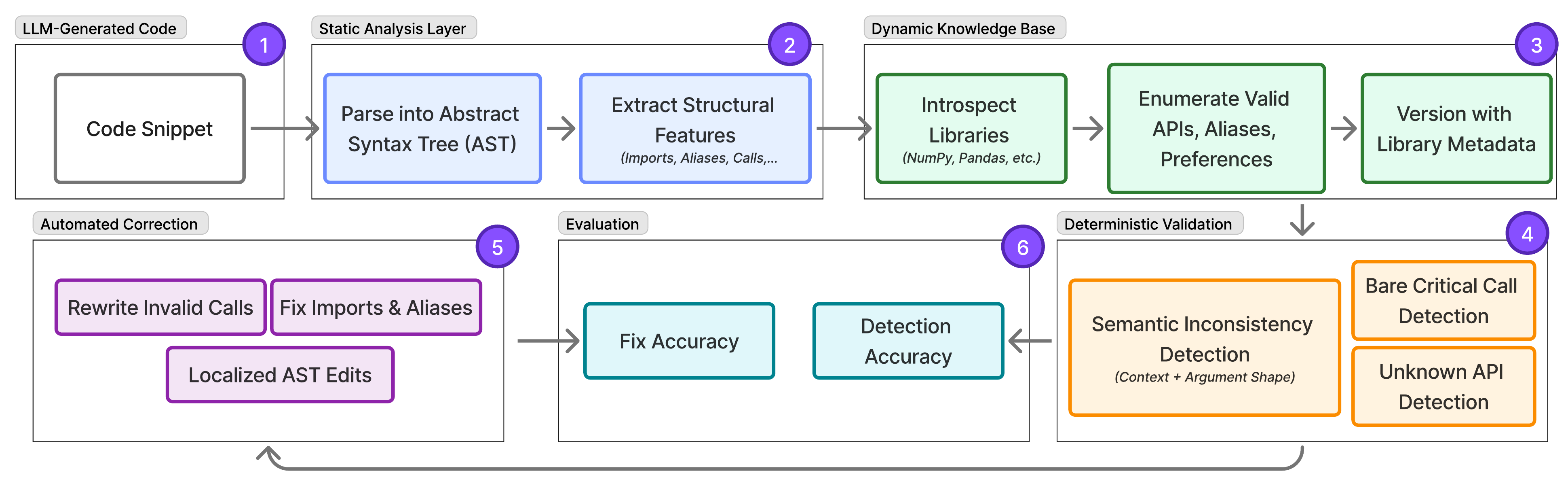}
    
    \caption{Hallucination Detection and Correction Framework}
    \label{fig:framework}
\end{figure*}

\subsection{Static Analysis Layer}
The framework starts with a code-generated snippet as illustrated in Figure \ref{fig:framework}-\circled{1}. The \textit{Static Analysis Layer} (\figref{fig:framework}-\circled{2}) parses the code snippet into an \AST. From this tree, we extract critical structural features: (i) all \texttt{import} statements and alias mappings (\eg \texttt{import pandas as pd}), (ii) fully qualified call sites (\eg \texttt{pd.\allowbreak read\_csv}), (iii) ``bare'' function calls lacking a module qualifier (\eg \texttt{loads}), and (iv) call arguments, especially string literals that imply intent (\eg \texttt{.csv}). This \AST-level analysis precisely identifies all references to libraries, functions, and objects, regardless of formatting.

\subsection{Dynamic Knowledge Base}
The imported import statements are fed into the \textbf{Dynamic Knowledge Base } (\figref{fig:framework}-\circled{3}). Our KB is not a fixed whitelist; it synthesizes knowledge via introspection. When encountering an \texttt{import pandas} statement, a builder utility imports the library, enumerates its public callables, and (for pandas) records common \texttt{DataFrame} and \texttt{Series} methods. The KB stores: (i) valid function/method names, (ii) common aliases (\texttt{np}, \texttt{pd}), and (iii) lightweight semantic preferences (\eg \texttt{requests.\allowbreak get} for retrieval). The KB is versioned with the library's \texttt{\_\_version\_\_} field to ensure reproducibility.

\subsection{Deterministic Validation}
The \textbf{Deterministic Validation}(\figref{fig:framework}-\circled{4}) layer validates each extracted function call against the populated KB. Issues are flagged in three categories:
\begin{enumerate}
    \item \textbf{Unknown API}: The callee (\eg \texttt{pd.\allowbreak read\_exel}) is missing from the KB. We suggest the closest match by edit distance.
    \item \textbf{Bare Critical Call}: A call (\eg \texttt{read\_csv}) appears without its required module alias. This is flagged for correction.
    \item \textbf{Semantic Inconsistency}: Syntactically valid code contradicts static cues. We use two cues: (a) \emph{Argument Shape}, where a \texttt{.csv} file extension mismatches \texttt{pd.\allowbreak read\_excel}, and (b) \emph{Intent Synonyms} (\eg "average" \(\Rightarrow\) \texttt{np.\allowbreak mean}).
\end{enumerate}

The validation process runs in $O(n \cdot m)$ time, where $n$ is the number of call sites and $m$ is the number of API entries, making it tractable even for large libraries.

\subsection{Automated Correction}

Identified issues are resolved through \textbf{Automated Correction}
(\figref{fig:framework}-\circled{5}), which applies localized \AST\ edits.
Misspellings are replaced with the closest valid symbol from the KB.
Contextual mismatches are rewritten (e.g.
\nolinkurl{pd.read_excel('f.csv')} \(\to\)
\nolinkurl{pd.read_csv('f.csv')}).
Bare calls trigger the insertion of missing imports with canonical aliases
(e.g. \texttt{import pandas as pd}) at the top of the file.
The modified \AST\ is then deterministically unparsed back to source code.

\subsection{Evaluation}
Finally, our framework is evaluated in the dataset from \S\ref{subsec:dataset} (\figref{fig:framework}-\circled{6}). We report two primary metrics:
\begin{enumerate}
    \item \textbf{Hallucination Detection Accuracy}: The detector's ability to identify \KCHs, treated as a binary classification problem, and reported using \textbf{Precision, Recall, and F1-score}.
    \item \textbf{Hallucination Fix Accuracy}: For snippets correctly identified as hallucinated, we measure the "fix accuracy," a binary score for whether our correction module produces functionally correct, runnable code.
\end{enumerate}
The entire detection and correction pipeline is highly efficient; in our experiments, the end-to-end analysis of all 200 samples completed in under 0.2 seconds on a single laptop CPU.

\subsection{Dataset Construction}
\label{subsec:dataset}
We construct our evaluation set of \textbf{200 total Python\footnote{The approach generalizes to any language with an AST representation and accessible reflection APIs, such as Java or TypeScript.} samples} by prompting GPT-5 with task-oriented instructions for five target libraries: \textit{numpy}, \textit{pandas}, \textit{requests}, \textit{matplotlib}, and \textit{json}. The dataset is composed of \textbf{161 hallucinated samples} and \textbf{39 clean (valid) samples}. The 161 hallucinated samples are curated examples of the \KCHs our system targets, falling into three main categories: (i) \textbf{Mis-typed API calls} (\eg \texttt{pd.\allowbreak read\_exel}), (ii) \textbf{Missing imports} (\eg \texttt{read\_csv}), and (iii) \textbf{Contextual mismatches} (\eg using \texttt{pd.\allowbreak read\_excel} to load a \texttt{.csv} file). The introspection-based design of the framework allows new libraries to be easily added.

\section{Results}
\label{sec:results}

We evaluated our framework on the curated 200-sample dataset described in \S\ref{subsec:dataset}. The results of this evaluation, divided into our two main tasks of detection and correction, show significant promise for our deterministic approach.

\begin{table}[htbp]

    \centering
    \caption{Confusion Matrix (Positive = Hallucinated)}
    \label{tab:confusion_matrix}
    \renewcommand{\arraystretch}{1.2} %
    \scalebox{0.8}{
    \begin{tabular}{l|cc|c}
\toprule
\multicolumn{1}{c|}{\multirow{2}{*}{\textbf{Actual}}} & \multicolumn{2}{c|}{\textbf{Predicted}}                              & \textbf{}             \\
\multicolumn{1}{c|}{}                                 & \textbf{Hallucinated}             & \textbf{Not Hallucinated}        & \textbf{Actual Total} \\ \hline
\textbf{Hallucinated}                                 & 141 \textcolor{gray}{\small (TP)} & 20 \textcolor{gray}{\small (FN)} & 161                   \\
\textbf{Not Hallucinated}                             & 0 \textcolor{gray}{\small (FP)}   & 39 \textcolor{gray}{\small (TN)} & 39                    \\ \hline
\textbf{Predicted Total}                              & 141                               & 59                               & 200                   \\ \hline
\end{tabular}
  
}
\end{table}

\begin{table}[htbp]
\centering
\caption{Hallucination Correction Performance (N=161)}
\label{tab:correction_performance}
\renewcommand{\arraystretch}{1.0}
\scalebox{0.93}{
\begin{tabular}{lrr}
\toprule
\textbf{Category} & \textbf{Count} & \textbf{Percentage} \\
\midrule
Total Hallucinated Samples (Identified) & 161 & 100.0\% \\
\quad \textit{Successfully Corrected} & \textit{124} & \textit{77.0\%} \\
\quad \textit{Uncorrected} & \textit{37} & \textit{23.0\%} \\
\bottomrule
\end{tabular}
}
\end{table}

\textbf{Detection Performance.} To answer our \ref{rq:explore}, we determine whether \KCHs could be \textit{detected} deterministically. The overall detection performance was strong, with 90\% accuracy and a 93.4\% F1-Score. The most critical finding is the framework's \textbf{100\% Precision}, which produced zero false positives and validated all 39 clean samples(Table \ref{tab:confusion_matrix}). This perfect precision confirms the reliability of the tool for developers. The overall \textbf{87.6\% Recall} reflects variance in our detailed analysis (Table \ref{tab:type_breakdown} and \ref{tab:library_breakdown}). Although detection was near-perfect for \texttt{Missing Imports} (97.9\%) and \texttt{numpy} (100.0\%), it was lowest for \texttt{Contextual Mismatches} (33.3\%) and \texttt{matplotlib.pyplot} (72.2\%), which account for our 20 False Negatives.

\vspace{-0.5em}
\begin{boxK}
\vspace{-0.5em}
    \textit{\ref{rq:explore}}
    Our framework detected KCHs with 100 \% precision, producing zero false positives and confirming its reliability for developers.
    \vspace{-1.5em}
\end{boxK}
\vspace{-0.2em}

\textbf{Correction Performance.} Our second research question (\ref{rq:beyond}) asked if these detected errors could be automatically \textit{corrected}. Our framework achieved a high overall \textbf{Fix Accuracy of 77.0\%} (Table \ref{tab:correction_performance}), successfully repairing 124 of the 161 hallucinated snippets. This success, however, varied significantly by \KCH type and library (Tables \ref{tab:type_breakdown} and \ref{tab:library_breakdown}). The fix rate was highest for \texttt{Missing Imports} (97.9\%) but dropped for \texttt{Mis-typed APIs} (70.0\%) and \texttt{pandas} (56.2\%), accounting for our 37 uncorrected samples. This high-level success still strongly suggests our deterministic approach is a viable strategy for handling the majority of common \KCHs. The high performance on \texttt{Missing Imports} and \texttt{Mis-typed APIs} is particularly notable, as it proves our framework's ability to resolve the most frequent and repetitive categories of \KCHs.

\vspace{-0.5em}
\begin{boxK}
\vspace{-0.5em}
    \textit{\ref{rq:beyond}}
    Our framework successfully and automatically corrected 77.0\% (124 of 161) of all identified hallucinations.
    \vspace{-1.0em}
\end{boxK}
\vspace{-0.2em}

\begin{table}[htbp]
\centering
\caption{Detection and Correction Performance by KCH Type}
\label{tab:type_breakdown}
\renewcommand{\arraystretch}{1.0}
\begin{tabularx}{\columnwidth}{@{} L c c c @{}} %
\toprule
\textbf{Hallucination Type} & \textbf{Smpl.} & \textbf{Detect. Rate} & \textbf{Corr. Acc.} \\
\midrule
Missing Imports & 48 & 97.9\% & 97.9\% \\
Mis-typed API Calls & 110 & 84.5\% & 70.0\% \\
Contextual Mismatches & 3 & 33.3\% & 0.0\% \\

\bottomrule
\end{tabularx}
\end{table}

\begin{table}[htbp]
\centering
\caption{Detection and Correction Performance by Library}
\label{tab:library_breakdown}
\renewcommand{\arraystretch}{1.0}
\begin{tabularx}{\columnwidth}{@{} L c c @{}} %
\toprule
\textbf{Library} & \textbf{Detect. Rate} & \textbf{Corr. Acc.} \\
\midrule
numpy & 100.0\% & 93.8\% \\
pandas & 85.1\% & 56.2\% \\
matplotlib.pyplot & 72.2\% & 65.5\% \\
json & 100.0\% & 81.2\% \\
requests & 100.0\% & 93.9\% \\
\bottomrule
\end{tabularx}
\end{table}

\section{Discussion and Future Work}
\label{sec:discussion}

Our results strongly support our hypothesis that a deterministic static framework can effectively resolve \KCHs. The 100\% precision in detection is the most critical finding, demonstrating that our framework operates without the ``alert fatigue'' of false positives. This result, combined with a 77.0\% automatic fix rate, positions our framework as a distinct and viable alternative to existing mitigations. Unlike prevention-based tools \cite{scholak-etal-2021-picard, poesia2022synchromeshreliablecodegeneration} that miss these syntactically-valid semantic errors, or non-deterministic ``LLM-in-the-loop'' repair \cite{blyth2025staticanalysisfeedbackloop, raviLLMLoop}, our method provides a reliable fix. Traditional static analyzers like mypy\cite{mypy_software} validate type correctness (\eg \textit{int + str})  using static stubs; they often fail to capture the deep semantic hallucinations inherent to LLMs (\ e.g., Does the API function \texttt{pd.read\_exel} actually exist in pandas?). Additionally, it extends post-processing AST approaches such as ``Structural Trimming'' \cite{zhang2025cutting} by focusing on resolution for correctness, not just deletion for safety.

A manual analysis of our 37 failed cases (20 false negatives, 17 failed corrections) reveals clear limitations. The false negatives were mostly mis-typed (\eg \texttt{plt.plotx} instead of \texttt{plt.plot}) or bare calls (\eg \texttt{plot(data, data)} instead of \texttt{plt.plot(data, data)}), concentrated in \texttt{matplotlib.pyplot}. The 17 failed corrections, worst in \texttt{pandas} (56.2\% accuracy), exposed a deeper challenge: our framework often corrected a \textit{surface-level typo} (\eg \texttt{np.arrya} $\to$ \texttt{np.array}) while missing the \textit{semantic error} (intent required \texttt{np.mean}). This confirms the limits of string matching and the need for semantic-intent analysis.

We must acknowledge the limitations of this study. Our 200-sample dataset, while manually curated, is not exhaustive, and its error distribution may not reflect real-world prevalence. The framework’s Knowledge Base was limited to five Python libraries, and our results do not guarantee generalizability. Our framework currently focuses on single-file, function-level analysis and does not yet handle multi-module dataflows or deep semantic intent inference; however, these are promising extensions rather than fundamental barriers. Finally, our approach deliberately targets KCHs and does not attempt to solve more complex, multi-line logical errors, which remain an important open challenge.

Our results, particularly the 100\% precision in detection, validate our deterministic static framework as a reliable foundation for practical developer tools. In the near term, we envision integrating this system as a lightweight \textit{semantic linter} within IDEs such as VS Code. As the LLM generates code, the analyzer would validate each call in real time, automatically correcting high-confidence hallucinations and flagging ambiguous ones for developer review. This hybrid human-in-the-loop approach could significantly improve trust and productivity.

To ensure generalizability beyond our initial study, we envision developing an automated ingestion pipeline. Instead of manual rule-writing, this pipeline uses structured scrapers and LLM-assisted parsers to extract AST-compatible signatures directly from official documentation and source code for any new library. This allows the KB to evolve continuously, maintaining accuracy across version updates (e.g., \texttt{pandas}~2.0 $\rightarrow$~3.0) and supporting large-scale CI/CD integration.

Together, these directions outline a practical evolution from a research prototype into an integrated toolchain for \textbf{trustworthy, self-correcting code generation}.

\section{Related Work}\label{sec:related_work}

Research into LLM-generated code errors has moved from initial taxonomies that identified bugs like "Hallucinated Objects" \cite{tambon2024bugslargelanguagemodels} to more specific analyzes of "Knowledge Conflicting Hallucinations" \cite{liu2024exploringevaluatinghallucinationsllmpowered} and definitions to test a novel repair strategy.

Current mitigation techniques fall into three categories, none of which fully solve the KCH problem. Prevention methods, such as PICARD \cite{scholak-etal-2021-picard} or Synchromesh \cite{poesia2022synchromeshreliablecodegeneration}, enforce syntactic correctness during generation but remain blind to semantic fact-conflicting errors. LLM-based repair uses a feedback loop, which is inherently non-deterministic \cite{blyth2025staticanalysisfeedbackloop, raviLLMLoop}. Finally, deletion-based repair like "Structural Trimming" \cite{zhang2025cutting} also uses an AST parser, but its goal is to prune the hallucinated code for safety. Our approach differs from all three: we propose a deterministic post-processing parser that focuses on resolution, using the AST to correct the hallucination, not just to prevent, delete, or ask for a new fix.

\section{Conclusions}\label{sec:conclusion}

This paper presents a deterministic, rule-based framework for detecting and repairing KCHs in LLM-generated code. Our empirical evaluation demonstrates that a non-probabilistic approach, leveraging AST analysis and a structured Knowledge Base, offers a highly reliable alternative to LLM-in-the-loop repair. By achieving 100\% precision in detection and an auto-correction rate of 77.0\%, we show that a significant class of semantic errors can be resolved without the risk of introducing new hallucinations. Ultimately, these results show that integrating static, deterministic logic into the generative pipeline is a viable path toward establishing the trust and structural integrity required for production-level AI-assisted software engineering.

\bibliographystyle{ACM-Reference-Format-num}
\bibliography{utils/references}


\begin{thebibliography}{14}

%%% the Web address.

\ifx \showCODEN    \undefined \def \showCODEN     #1{\unskip}     \fi
\ifx \showDOI      \undefined \def \showDOI       #1{#1}\fi
\ifx \showISBNx    \undefined \def \showISBNx     #1{\unskip}     \fi
\ifx \showISBNxiii \undefined \def \showISBNxiii  #1{\unskip}     \fi
\ifx \showISSN     \undefined \def \showISSN      #1{\unskip}     \fi
\ifx \showLCCN     \undefined \def \showLCCN      #1{\unskip}     \fi
\ifx \shownote     \undefined \def \shownote      #1{#1}          \fi
\ifx \showarticletitle \undefined \def \showarticletitle #1{#1}   \fi
\ifx \showURL      \undefined \def \showURL       {\relax}        \fi
% invisible to TeX
\providecommand\bibfield[2]{#2}
\providecommand\bibinfo[2]{#2}
\providecommand\natexlab[1]{#1}
\providecommand\showeprint[2][]{arXiv:#2}

\bibitem[\protect\citeauthoryear{Blyth, Licorish, Treude, and Wagner}{Blyth et~al\mbox{.}}{2025}]%
        {blyth2025staticanalysisfeedbackloop}
\bibfield{author}{\bibinfo{person}{Scott Blyth}, \bibinfo{person}{Sherlock~A. Licorish}, \bibinfo{person}{Christoph Treude}, {and} \bibinfo{person}{Markus Wagner}.} \bibinfo{year}{2025}\natexlab{}.
\newblock \bibinfo{title}{Static Analysis as a Feedback Loop: Enhancing LLM-Generated Code Beyond Correctness}.
\newblock
\newblock
\showeprint[arxiv]{cs.SE/2508.14419}
\urldef\tempurl%
\url{https://arxiv.org/abs/2508.14419}
\showURL{%
\tempurl}


\bibitem[\protect\citeauthoryear{Khati, Liu, Palacio, Zhang, and Poshyvanyk}{Khati et~al\mbox{.}}{2025}]%
        {khatiTrust}
\bibfield{author}{\bibinfo{person}{Dipin Khati}, \bibinfo{person}{Yijin Liu}, \bibinfo{person}{David~N. Palacio}, \bibinfo{person}{Yixuan Zhang}, {and} \bibinfo{person}{Denys Poshyvanyk}.} \bibinfo{year}{2025}\natexlab{}.
\newblock \showarticletitle{Mapping the Trust Terrain: LLMs in Software Engineering - Insights and Perspectives}.
\newblock \bibinfo{journal}{\emph{ACM Trans. Softw. Eng. Methodol.}} (\bibinfo{date}{Oct.} \bibinfo{year}{2025}).
\newblock
\showISSN{1049-331X}
\urldef\tempurl%
\url{https://doi.org/10.1145/3771282}
\showDOI{\tempurl}


\bibitem[\protect\citeauthoryear{Lab}{Lab}{2025}]%
        {repo}
\bibfield{author}{\bibinfo{person}{SEMERU Lab}.} \bibinfo{year}{2025}\natexlab{}.
\newblock \bibinfo{title}{Hallucinations-in-code}.
\newblock \bibinfo{howpublished}{\url{https://github.com/WM-SEMERU/Hallucinations-in-Code}}.
\newblock
\newblock
\shownote{Accessed: 2025-10-19.}


\bibitem[\protect\citeauthoryear{Lee, Song, Kim, Kim, Kim, et~al\mbox{.}}{Lee et~al\mbox{.}}{2025}]%
        {lee2025hallucinationcodegenerationllms}
\bibfield{author}{\bibinfo{person}{Yunseo Lee}, \bibinfo{person}{John~Youngeun Song}, \bibinfo{person}{Dongsun Kim}, \bibinfo{person}{Jindae Kim}, \bibinfo{person}{Mijung Kim}, {et~al\mbox{.}}} \bibinfo{year}{2025}\natexlab{}.
\newblock \bibinfo{title}{Hallucination by Code Generation LLMs: Taxonomy, Benchmarks, Mitigation, and Challenges}.
\newblock
\newblock
\showeprint[arxiv]{cs.SE/2504.20799}
\urldef\tempurl%
\url{https://arxiv.org/abs/2504.20799}
\showURL{%
\tempurl}


\bibitem[\protect\citeauthoryear{Lehtosalo et~al\mbox{.}}{Lehtosalo et~al\mbox{.}}{2025}]%
        {mypy_software}
\bibfield{author}{\bibinfo{person}{Jukka Lehtosalo} {et~al\mbox{.}}} \bibinfo{year}{2025}\natexlab{}.
\newblock \bibinfo{title}{{python/mypy: Optional static typing for Python}}.
\newblock
\newblock
\urldef\tempurl%
\url{https://github.com/python/mypy}
\showURL{%
\tempurl}


\bibitem[\protect\citeauthoryear{Liu, Liu, Shi, Huang, Wang, et~al\mbox{.}}{Liu et~al\mbox{.}}{2024}]%
        {liu2024exploringevaluatinghallucinationsllmpowered}
\bibfield{author}{\bibinfo{person}{Fang Liu}, \bibinfo{person}{Yang Liu}, \bibinfo{person}{Lin Shi}, \bibinfo{person}{Houkun Huang}, \bibinfo{person}{Ruifeng Wang}, {et~al\mbox{.}}} \bibinfo{year}{2024}\natexlab{}.
\newblock \bibinfo{title}{Exploring and Evaluating Hallucinations in LLM-Powered Code Generation}.
\newblock
\newblock
\showeprint[arxiv]{cs.SE/2404.00971}
\urldef\tempurl%
\url{https://arxiv.org/abs/2404.00971}
\showURL{%
\tempurl}


\bibitem[\protect\citeauthoryear{Peng, Kalliamvakou, Cihon, and Demirer}{Peng et~al\mbox{.}}{2023}]%
        {peng2023impactaideveloperproductivity}
\bibfield{author}{\bibinfo{person}{Sida Peng}, \bibinfo{person}{Eirini Kalliamvakou}, \bibinfo{person}{Peter Cihon}, {and} \bibinfo{person}{Mert Demirer}.} \bibinfo{year}{2023}\natexlab{}.
\newblock \bibinfo{title}{The Impact of AI on Developer Productivity: Evidence from GitHub Copilot}.
\newblock
\newblock
\showeprint[arxiv]{cs.SE/2302.06590}
\urldef\tempurl%
\url{https://arxiv.org/abs/2302.06590}
\showURL{%
\tempurl}


\bibitem[\protect\citeauthoryear{Poesia, Polozov, Le, Tiwari, Soares, et~al\mbox{.}}{Poesia et~al\mbox{.}}{2022}]%
        {poesia2022synchromeshreliablecodegeneration}
\bibfield{author}{\bibinfo{person}{Gabriel Poesia}, \bibinfo{person}{Oleksandr Polozov}, \bibinfo{person}{Vu Le}, \bibinfo{person}{Ashish Tiwari}, \bibinfo{person}{Gustavo Soares}, {et~al\mbox{.}}} \bibinfo{year}{2022}\natexlab{}.
\newblock \bibinfo{title}{Synchromesh: Reliable code generation from pre-trained language models}.
\newblock
\newblock
\showeprint[arxiv]{cs.LG/2201.11227}
\urldef\tempurl%
\url{https://arxiv.org/abs/2201.11227}
\showURL{%
\tempurl}


\bibitem[\protect\citeauthoryear{Ravi, Bradshaw, Ruberto, Jahangirova, and Terragni}{Ravi et~al\mbox{.}}{2025}]%
        {raviLLMLoop}
\bibfield{author}{\bibinfo{person}{Ravin Ravi}, \bibinfo{person}{Dylan Bradshaw}, \bibinfo{person}{Stefano Ruberto}, \bibinfo{person}{Gunel Jahangirova}, {and} \bibinfo{person}{Valerio Terragni}.} \bibinfo{year}{2025}\natexlab{}.
\newblock \showarticletitle{LLMLOOP: Improving LLM-Generated Code and Tests through Automated Iterative Feedback Loops}.
\newblock
\urldef\tempurl%
\url{https://doi.org/10.1109/ICSME64153.2025.00109}
\showDOI{\tempurl}


\bibitem[\protect\citeauthoryear{Scholak, Schucher, and Bahdanau}{Scholak et~al\mbox{.}}{2021}]%
        {scholak-etal-2021-picard}
\bibfield{author}{\bibinfo{person}{Torsten Scholak}, \bibinfo{person}{Nathan Schucher}, {and} \bibinfo{person}{Dzmitry Bahdanau}.} \bibinfo{year}{2021}\natexlab{}.
\newblock \showarticletitle{{PICARD}: Parsing Incrementally for Constrained Auto-Regressive Decoding from Language Models}. In \bibinfo{booktitle}{\emph{Proceedings of the 2021 Conference on Empirical Methods in Natural Language Processing}}, \bibfield{editor}{\bibinfo{person}{Marie-Francine Moens}, \bibinfo{person}{Xuanjing Huang}, \bibinfo{person}{Lucia Specia}, {and} \bibinfo{person}{Scott Wen-tau Yih}} (Eds.). \bibinfo{publisher}{Association for Computational Linguistics}, \bibinfo{address}{Online and Punta Cana, Dominican Republic}, \bibinfo{pages}{9895--9901}.
\newblock
\urldef\tempurl%
\url{https://doi.org/10.18653/v1/2021.emnlp-main.779}
\showDOI{\tempurl}


\bibitem[\protect\citeauthoryear{Tambon, Dakhel, Nikanjam, Khomh, Desmarais, et~al\mbox{.}}{Tambon et~al\mbox{.}}{2024}]%
        {tambon2024bugslargelanguagemodels}
\bibfield{author}{\bibinfo{person}{Florian Tambon}, \bibinfo{person}{Arghavan~Moradi Dakhel}, \bibinfo{person}{Amin Nikanjam}, \bibinfo{person}{Foutse Khomh}, \bibinfo{person}{Michel~C. Desmarais}, {et~al\mbox{.}}} \bibinfo{year}{2024}\natexlab{}.
\newblock \bibinfo{title}{Bugs in Large Language Models Generated Code: An Empirical Study}.
\newblock
\newblock
\showeprint[arxiv]{cs.SE/2403.08937}
\urldef\tempurl%
\url{https://arxiv.org/abs/2403.08937}
\showURL{%
\tempurl}


\bibitem[\protect\citeauthoryear{Wang, Zhou, Song, Huang, Chen, et~al\mbox{.}}{Wang et~al\mbox{.}}{2025}]%
        {wang2025understandingcharacteristicscodegeneration}
\bibfield{author}{\bibinfo{person}{Zhijie Wang}, \bibinfo{person}{Zijie Zhou}, \bibinfo{person}{Da Song}, \bibinfo{person}{Yuheng Huang}, \bibinfo{person}{Shengmai Chen}, {et~al\mbox{.}}} \bibinfo{year}{2025}\natexlab{}.
\newblock \bibinfo{title}{Towards Understanding the Characteristics of Code Generation Errors Made by Large Language Models}.
\newblock
\newblock
\showeprint[arxiv]{cs.SE/2406.08731}
\urldef\tempurl%
\url{https://arxiv.org/abs/2406.08731}
\showURL{%
\tempurl}


\bibitem[\protect\citeauthoryear{Wen, Zhu, Liu, Ren, Du, et~al\mbox{.}}{Wen et~al\mbox{.}}{2025}]%
        {wen2025fixingfunctionlevelcodegeneration}
\bibfield{author}{\bibinfo{person}{Hao Wen}, \bibinfo{person}{Yueheng Zhu}, \bibinfo{person}{Chao Liu}, \bibinfo{person}{Xiaoxue Ren}, \bibinfo{person}{Weiwei Du}, {et~al\mbox{.}}} \bibinfo{year}{2025}\natexlab{}.
\newblock \bibinfo{title}{Fixing Function-Level Code Generation Errors for Foundation Large Language Models}.
\newblock
\newblock
\showeprint[arxiv]{cs.SE/2409.00676}
\urldef\tempurl%
\url{https://arxiv.org/abs/2409.00676}
\showURL{%
\tempurl}


\bibitem[\protect\citeauthoryear{Zhang}{Zhang}{2025}]%
        {zhang2025cutting}
\bibfield{author}{\bibinfo{person}{Yage Zhang}.} \bibinfo{year}{2025}\natexlab{}.
\newblock \showarticletitle{Cutting the Root of Hallucination: Structural Trimming for Vulnerability Mitigation in Code {LLM}s}. In \bibinfo{booktitle}{\emph{Second Conference on Language Modeling}}.
\newblock
\urldef\tempurl%
\url{https://openreview.net/forum?id=dU4Y2sNfJ2}
\showURL{%
\tempurl}


\end{thebibliography}

\end{document}